\documentclass[11pt,dvips]{article}

\usepackage{amsmath,amssymb,amsthm,amscd,latexsym}
\usepackage[toc,title,titletoc]{appendix}
\usepackage{cite}
\usepackage{graphicx,epsfig}
\usepackage[breaklinks,
colorlinks=true,
linkcolor=red,
urlcolor=blue]{hyperref}
\usepackage{breakurl}

\RequirePackage{color}

\allowdisplaybreaks[1]

\def\nk{n_{\rm b}}

\def\rfr#1{eq. (\ref{#1})}

\def\eqi{\begin{equation}}
\def\eqf{\end{equation}}
\def\eqia{\begin{eqnarray}}
\def\eqfa{\end{eqnarray}}

\def\rp#1#2{{#1\over#2}}
\def\lb#1{\label{#1}}

\def\bds#1{\boldsymbol{#1}}

\begin{document}

\title{A flyby anomaly for Juno? Not from standard physics}

\author{L. Iorio\\ Ministero dell'Istruzione, dell'Universit$\grave{\textrm{a}}$ e della Ricerca (M.I.U.R.)-Istruzione \\ Fellow of the Royal Astronomical Society (F.R.A.S.)\\ Viale Unit$\grave{\textrm{a}}$ di Italia 68, 70125, Bari (BA), Italy\\ Tel. +39 329 2399167. Fax: N/A. email: lorenzo.iorio@libero.it}

\maketitle

\begin{abstract}
 An empirical formula recently appeared in the literature to explain the observed anomalies of about  $\Delta\dot\rho\approx 1-10$ mm s$^{-1}$ in the geocentric range-rates $\dot\rho$ of the Galileo, NEAR and Rosetta spacecraft at some of their past perigee passages along unbound, hyperbolic trajectories. It  predicts an anomaly of the order of $6$ mm s$^{-1}$ for the recent flyby of Juno, occurred on 9 October 2013. Data analyses to confirm or disproof it are currently ongoing.
 We numerically calculate the impact on the geocentric Juno's range rate of some classical and general relativistic dynamical effects which are either unmodelled or mismodelled to a certain level in the software used to process the data. They are: a) The first even zonal harmonic coefficient $J_2$ of the multipolar expansion of the terrestrial gravitational potential causing orbital perturbations both at the $\left.{\rm a}^{'}\right)$ Newtonian ($J_2$) and at the $\left. {\rm a}^{''}\right)$ first post-Newtonian level ($J_2 c^{-2}$) b) The post-Newtonian gravitoelectric (GE) Schwarschild-like component of the Earth's gravitational field c) The post-Newtonian gravitomagnetic (GM) Lense-Thirring effect. The magnitudes of their mismodeled and nominal range-rate signatures are: $\left. {\rm a}^{'}\right)$ $\Delta\dot\rho_{\sigma_{J_2}} \approx 1$ $\mu$m s$^{-1}$ $\left. {\rm a}^{''}\right)$ $\Delta\dot\rho_{J_2 c^{-2}} \approx 0.015$ $\mu$m s$^{-1}$ b) $\Delta\dot\rho_{\rm GE} \approx 25$ $\mu$m s$^{-1}$
 c) $\Delta\dot\rho_{\rm GM} \approx 0.05$ $\mu$m s$^{-1}$. If a flyby anomaly as large as a few mm s$^{-1}$ will be finally found also for Juno, it will not be due to any of these standard gravitational effects.
 It turns out that a Rindler-type radial extra-acceleration of the same magnitude as in the Pioneer anomaly would impact the Juno's range-rate at a $\Delta\dot\rho_{\rm Rin} \approx 1.5$ $\mu$m s$^{-1}$ level. Regardless of the quest for the flyby anomaly, all such effects are undetectable.
\end{abstract}

\leftline{Keywords:}
\leftline{Experimental studies of gravity;}
\leftline{Experimental tests of gravitational theories;}
\leftline{Modified theories of gravity;}
\leftline{Lunar, planetary, and deep-space probes}

\leftline
{PACS:}
\leftline{04.80.-y;}
\leftline{04.80.Cc;}
\leftline{04.50.Kd;}
\leftline{95.55.Pe}



\section{Introduction}\lb{Introduzione}
On 9 October 2013, the NASA's spacecraft Juno\footnote{See also \url{http://missionjuno.swri.edu/} and \url{http://www.nasa.gov/mission_pages/juno/} on the Internet.} \cite{2007AcAau..61..932M} made an Earth flyby passing to within  561 km of our planet at $19:21$ GMT to gain the required gravitational energy to reach Jupiter, its final target, in July 2016 \cite{2011Icar..216..440H}.

Such an event raised interest \cite{Clark013, Scuka013, 2013AGUFMSM33B2187A, 2013arXiv1312.1139B} because of its potential capability to shed more light on one of the recently reported astrometric anomalies in the Solar System \cite{2010IAUS..261..189A}: the so-called flyby anomaly \cite{2007NewA...12..383A, 2008mgm..conf.2564L, 2008PhRvL.100i1102A, 2009arXiv0910.1321N, 2009SSRv..148..169T}. It consists of a small, unexpected increase of the geocentric range-rate \eqi\Delta\dot \rho\approx 1-10\ {\rm mm\ s}^{-1}\eqf experienced by some spacecraft (Galileo, NEAR, Rosetta) approaching the Earth along unbound, hyperbolic trajectories in occasion of some of their flybys. At present, no satisfactory explanations exist for such a phenomenon in terms of both conventional gravitational and non-gravitational physics; see, e.g., \cite{2008ASSL..349...75L, 2009SSRv..148..169T} and references therein. In particular, in \cite{2011AnP...523..439R} it was shown that the thermal effects which should be responsible for most of the Pioneer anomaly \cite{2010LRR....13....4T} could not explain the Rosetta flyby anomaly. Possible spacecraft electrostatic charging effects in terms of a Lorentz force were ruled out in \cite{2010JGCD...33.1115A}. For the\textcolor{black}{-negligible-}impact of the general relativistic gravitomagnetic Lense-Thirring effect on the motion of a test particle in hyperbolic motion, see \cite{2009ScReE2009.7695I, 2010cosp...38.3845H}. \textcolor{black}{Another negative result in term of the Kerr geometry in the context of Conformal Gravity was recently obtained in \cite{2014arXiv1401.6503V}}. Several more or less sound explanations in terms of non-conventional physics have been put forth so far \cite{2007arXiv0711.2781B, 2008arXiv0803.1370N, 2008arXiv0804.0039C, 2008arXiv0804.2198S, 2008arXiv0806.0334P, 2008arXiv0807.3158G, 2008arXiv0809.1888M, 2009AIPC.1103..226L, 2009AIPC.1103..302M, 2009AIPC.1103..311F, 2009arXiv0904.0383H, 2009arXiv0909.5150P, 2010arXiv1006.3555B, 2010IJMPA..25..815C, 2010arXiv1004.0826M, 2011IJMPE..20...78L, 2011arXiv1102.2945R, 2011arXiv1105.3857H, 2011arXiv1109.0256P, 2012AIPC.1483..260R, 2012PrPh....2...39T, 2014arXiv1404.1101P, Acedo2014} with mixed success. We mention also a proposed modification of inertia  \cite{2008JBIS...61..373M, 2008MNRAS.389L..57M}, and the effect of Earth-bound Dark Matter \cite{2009PhRvD..79b3505A, 2009arXiv0910.1564A, 2010IJMPA..25.4577A, 2013IJMPA..2850074A}. Proposals have been made to test the flyby anomaly with dedicated future space-based missions \cite{2011arXiv1109.2779B, 2012IJMPD..2150035B, 2013P&SS...79...76P}.

The opportunity offered by Juno is, in principle, interesting also because of the relatively low altitude  of its flyby of Earth. The expected effect is of the order of \textcolor{black}{\cite{2013AGUFMSM33B2187A}\footnote{\textcolor{black}{It may interesting to note that  an anomaly with the same magnitude but with the opposite sign was predicted in \cite{2013arXiv1312.1139B} .}}} \eqi \Delta\dot\rho_{\rm Juno} \approx \textcolor{black}{7}\ {\rm mm\ s}^{-1}.\lb{flyby}\eqf  The figure in \rfr{flyby} can be obtained\textcolor{black}{, e.g.,} by using the empirical formula devised in \cite{2008PhRvL.100i1102A} to accommodate \textcolor{black}{some of} the previously observed flybys of the other spacecraft. \textcolor{black}{However, it should be recalled that the formula by Anderson et al. \cite{2008PhRvL.100i1102A} gives wrong (not null) anomaly predictions for the second and third Rosetta flybys}. In waiting for the final outcome of the ongoing  analysis by NASA/JPL of the data collected by ESA aimed to establish  if the flyby anomaly exists also for Juno or not, in this paper we will quantitatively look at the effects of some standard Newtonian/Einsteinian gravitational effects on the geocentric range-rate of the Jupiter-targeted spacecraft at the epoch of its terrestrial flyby. Some of them, like the Lense-Thirring effect, recently detected in the Earth's gravitational field with a claimed $19\%$ accuracy \cite{2011PhRvL.106v1101E}, are unmodeled in the softwares used to process the spacecraft's data, while others are modeled with a necessarily limited accuracy. Our aim is to calculate the size of such range-rate signals to see if they are relevant at a $\sigma_{\dot\rho}\approx$ mm s$^{-1}$ level of accuracy and, in particular, if they could allow for an effect as large as \rfr{flyby}.
\section{Numerical simulations}\lb{numerico}
In order to investigate the impact of some gravitational effects which, in principle, may induce a flyby anomaly for Juno, we numerically integrate its equations of motion  in a geocentric reference frame with Cartesian orthogonal coordinates. For each additional acceleration ${\bds A}_{\rm pert}$ with respect to the Newtonian monopole ${\bds A}_{\rm N}$, viewed as a small perturbation of it, we perform two numerical integrations: one in which the total acceleration is ${\bds A}_{\rm tot} = {\bds A}_{\rm N} + {\bds A}_{\rm pert}$, and one in which we keep only ${\bds A}_{\rm N}$. Both the integrations share the same initial conditions, retrieved from the HORIZONS WEB interface by NASA/JPL. Then, from the resulting time series $\{x_{\rm pert}(t),y_{\rm pert}(t),z_{\rm pert}(t)\}$ and $\{x_{\rm N}(t),y_{\rm N}(t),z_{\rm N}(t)\}$ for the geocentric coordinates we produce two time series $\dot\rho_{\rm pert}(t)$ and $\rho_{\rm N}(t)$ for the range-rate $\rho$ and take, their difference to obtain $\Delta\dot\rho$ which singles out the expected signature of the effect one is interested in on the Juno's range rate. The integration time span is $\Delta t = 4000$ s, starting from the shadow entry, so that the flyby occurs after $1260$ s from $t_0=0$.

In Figure \ref{curve} we depicts our results for the following dynamical features.
\begin{itemize}
\item The Newtonian effect of the Earth's oblateness, parameterized by the first even zonal harmonic $J_2 = -\sqrt{5}\ {\overline{C}}_{2,0}$, where ${\overline{C}}_{\ell,m}$ are the normalized  Stokes coefficients of degree $\ell$ and order $m$ of the geopotential \cite{Heis67}. In the left upper corner, its nominal range-rate shift is depicted. It shows a peak-to-peak amplitude of a few m s$^{-1}$. Actually, global Earth's gravity field models are usually adopted in the data reduction softwares, so that one has to look just at the residual range-rate signature left by the unavoidable mismodeling in $J_2$ as a potential cause for a flyby anomaly. By conservatively evaluating $\sigma_{{\overline{C}}_{2,0}}$ as in \cite{2012JHEP...05..073I} on the basis of the independent approach recently put forth in \cite{2012JGeod..86...99W}, we obtained the signal displayed in the right upper corner of Figure \ref{curve}. Its amplitude is as little as 1 $\mu$m s$^{-1}$.
\item The first post-Newtonian (1PN) gravitoelectric (GE), Schwarzschild-like component of the Earth's field, usually modeled in the data analysis softwares. The left mid panel shows its nominal range-rate signature, which amounts to 25 $\mu$m s$^{-1}$.
\item     The first post-Newtonian (1PN) gravitomagnetic (GM), Lense-Thirring component of the Earth's field, usually unmodeled in the data analysis softwares. The peak-to-peak amplitude of its signal, shown in the right mid panel of Figure \ref{curve}, is as little as $0.05$ $\mu$m s$^{-1}$.
\item     The first post-Newtonian (1PN) aspherical component of the Earth's field, proportional to $J_2 c^{-2}$
\cite{1988CeMec..42...81S, 1990CeMDA..47..205H, 1991ercm.book.....B}, usually unmodeled in the data analysis softwares. Its effect on the Juno's range-rate, displayed in the left lower corner of Figure \ref{curve},  amounts to $0.015$ $\mu$m s$^{-1}$.
\item A Rindler-type radial acceleration \cite{2010PhRvL.105u1303G, 2011PhRvD..83l4024C, 2011IJMPD..20.2761G} with the same magnitude as in the Pioneer anomaly (right lower corner). As elucidated in \cite{2011PhRvD..83l4024C, 2011IJMPD..20.2761G}, it may not be applicable to huge bodies of astronomical size, contrary to man-made objects such as  spacecraft like, e.g., the Pioneer probes. Indeed for a body of mass $m_{\rm b}$ and size $d_{\rm b}$, the condition \cite{2011PhRvD..83l4024C, 2011IJMPD..20.2761G}
    \eqi \rp{Gm_{\rm b}}{d_{\rm b}}\lesssim  |A_{\rm Rin}| r\lb{condizione}\eqf
    must be satisfied. Since\footnote{We take the largest dimension of the solar arrays for $d_{\rm J}$.} \cite{2007AcAau..61..932M} $m_{\rm J} =3625 $ kg, $d_{\rm J}\approx 9$ m, and $r=6894$ km at the flyby, \rfr{condizione} is fully satisfied for Juno. Indeed, $Gm_{\rm J} d_{\rm J}^{-1}= 2.7\times 10^{-8}$ m$^2$ s$^{-2}$, while $|A_{\rm Rin}|r_{\rm J}=6\times 10^{-3}$ m$^2$ s$^{-2}$.
     The magnitude of the putative Rindler-type effect on the Juno's range-rate turns out to be $1.5$ $\mu$m s$^{-1}$.
\end{itemize}
\begin{figure*}
\centering
\begin{tabular}{cc}
\epsfig{file=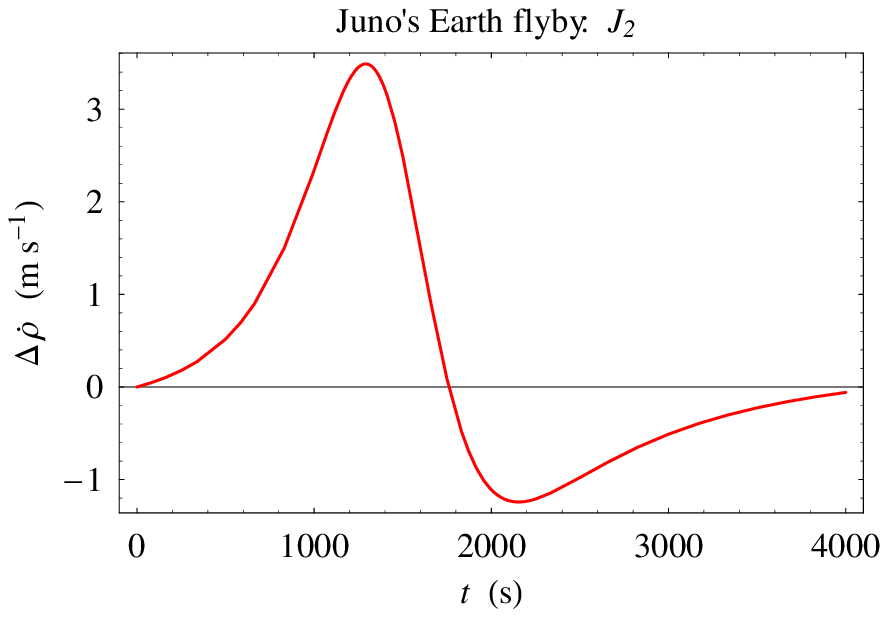,width=0.40\linewidth,clip=} & \epsfig{file=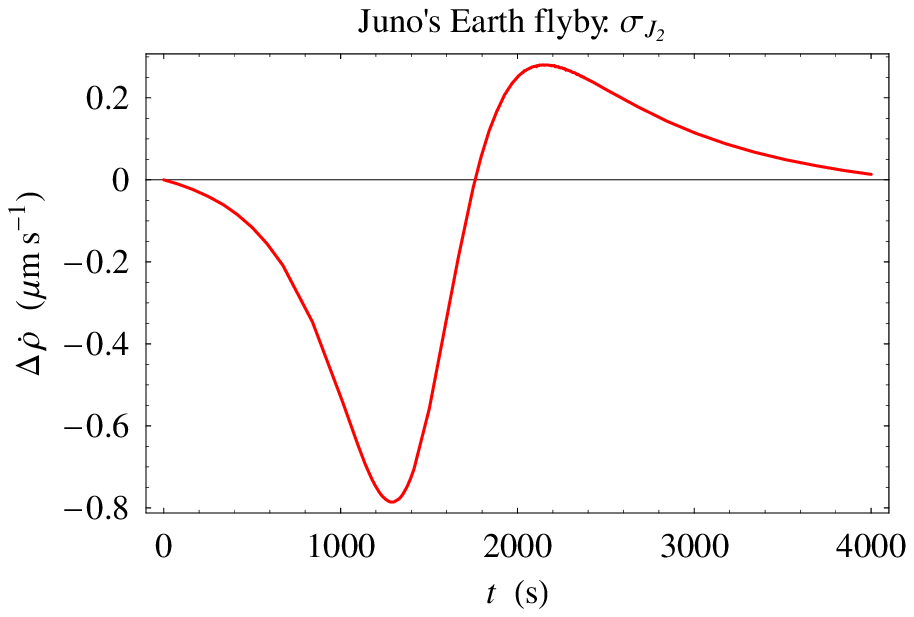,width=0.40\linewidth,clip=}\\
\epsfig{file=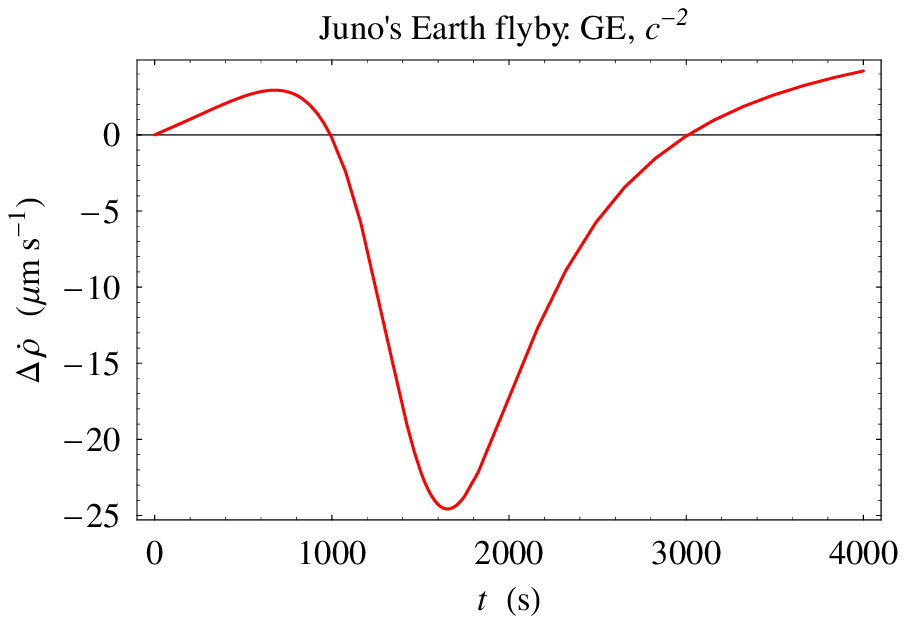,width=0.40\linewidth,clip=} & \epsfig{file=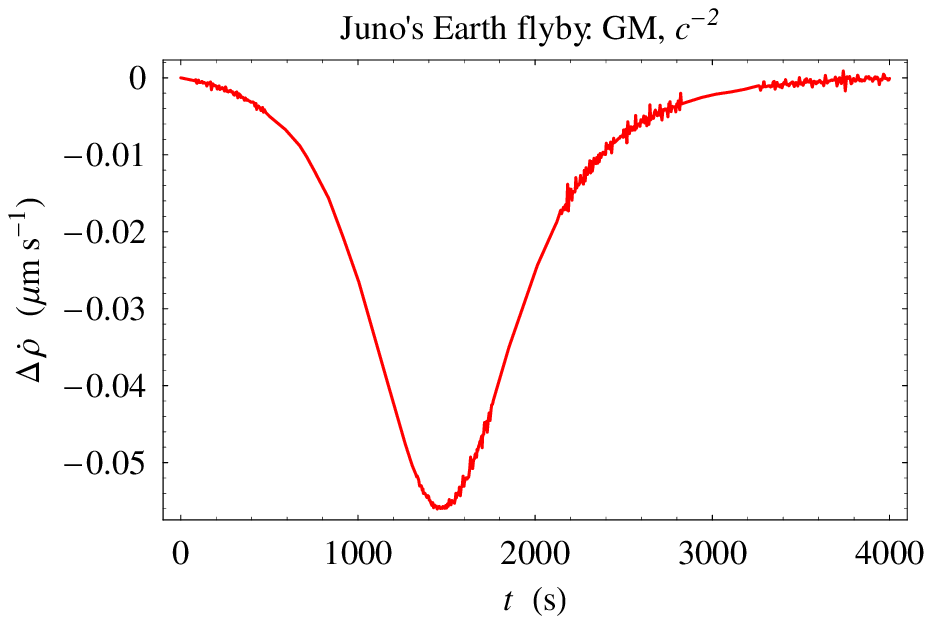,width=0.40\linewidth,clip=}\\
\epsfig{file=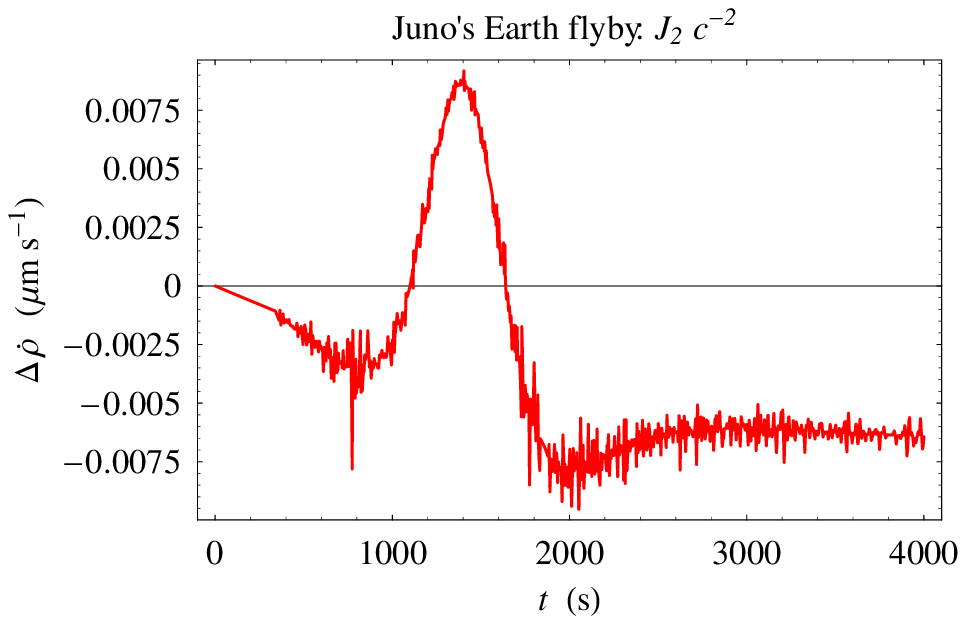,width=0.40\linewidth,clip=} & \epsfig{file=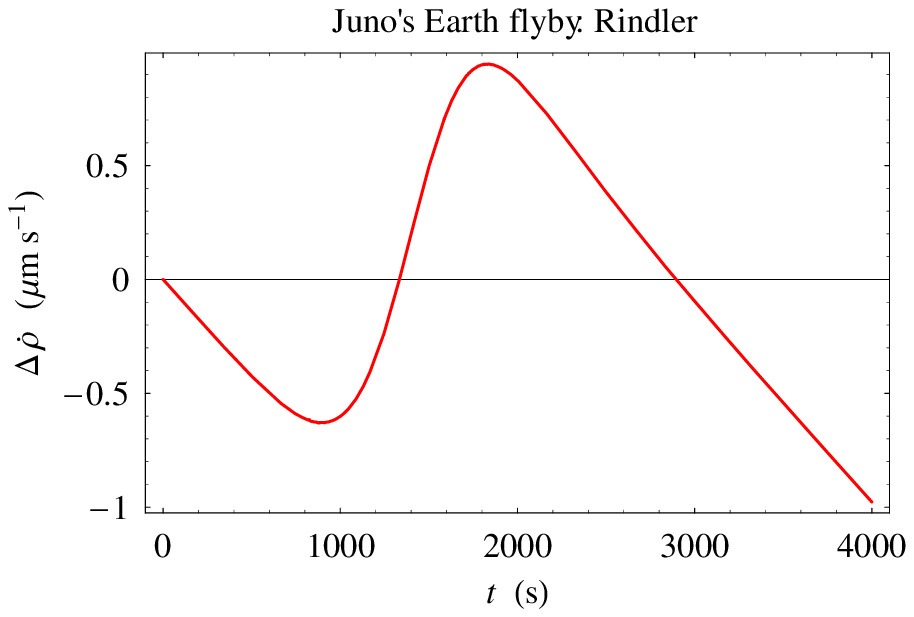,width=0.40\linewidth,clip=}\\
\end{tabular}
\caption{Simulated geocentric range-rate signatures of Juno at the Earth's flyby induced by various standard and non-standard dynamical effects. The units are m s$^{-1}$ for the nominal $J_2$ signal in the upper left corner, while  $\mu$m s$^{-1}$ are used for all the other effects. For the signal induced by the mismodeling in the first even zonal harmonic of the geopotential, displayed  in the upper right corner, the conservative value $\sigma_{{\overline{C}}_{2,0}} = 1.09\times 10^{-10}$  \cite{2012JHEP...05..073I}, calculated with the approach in \cite{2012JGeod..86...99W}, is adopted. The Rindler-type curve, placed in the lower right corner, is obtained with the value $A_{\rm Rin} = 8.7\times 10^{-10}$ m s$^{-2}$. }\lb{curve}
\end{figure*}
It can be noticed that none of the effects considered is able to impact the Juno's range-rate at a mm s$^{-1}$ level. Moreover, the smallness of the  1PN signatures which will likely not be modeled in the data analysis (GM + $J_2 c^{-2}$)  should make them practically undetectable. The same holds also for a Rindler-type acceleration $A_{\rm Rin}$ large enough to explain the Pioneer anomaly. On the one hand, it could not be independently tested with the Moon's motion. Indeed, $Gm_{\rm M} d_{\rm M}^{-1}=2.8\times 10^6$ m$^2$ s$^{-2}$, $|A_{\rm Rin}|r_{\rm M} = 3.3\times 10^{-1}$ m$^2$ s$^{-2}$, so that the condition of  \rfr{condizione} would not be satisfied for the Moon. On the other hand, $A_{\rm Rin}$ could, in principle, certainly affect all the artificial satellites orbiting the Earth along bound trajectories. For, e.g., LARES, it is \cite{2011AcAau..69..127P, 2013AcAau..91..313P} $m_{\rm LR}= 386.8$ kg, $d_{\rm LR} = 36.4$ cm, and $r_{\rm LR}=7828$ km; thus $Gm_{\rm LR}d_{\rm LR}^{-1}= 7\times 10^{-8}$ m$^2$ s$^{-2}$, $|A_{\rm Rin}|r_{\rm LR} = 6.8\times 10^{-3}$ m$^2$ s$^{-2}$, and the condition of \rfr{condizione} is fully satisfied. From \cite{2013NewA...23...63R}, it turns out that the perigee precession of LARES is constrained down to $\sigma_{\dot\omega_{\rm LR}} = 522$ milliarcseconds per year (mas yr$^{-1}$)  level because of unmodeled empirical accelerations in the along-track direction. This rules out the possible existence of an anomalous Rindler-type acceleration for the Earth as large as $A_{\rm Rin}= 8.7\times 10^{-10}$ m s$^{-2}$ since it would induce an anomalous perigee precession\footnote{The pericenter precession caused by a radial uniform extra-acceleration $A$ is \cite{2006NewA...11..600I, 2006MNRAS.370.1519S, 2006MNRAS.371..626S} $\dot\omega = A\sqrt{1-e^2} \nk^{-1} a^{-1}$, where $\nk,a,e$ are the satellite's mean motion, semimajor axis, and eccentricity, respectively.} for LARES  as large as $\dot\omega_{\rm Rin} = 800$ mas yr$^{-1}$. From the point of view of the Juno's flyby anomaly, an even smaller magnitude of a putative $A_{\rm Rin}$, compatible with the LARES bound, would be even more insignificant.
\section{Conclusions}\lb{conclusione}
After its flyby of Earth occurred on 9 October 2013, analyses of the data collected by ESA have been started by NASA/JPL  to determine if also Juno, now en route to Jupiter, will exhibit the so-called flyby anomaly which was detected in some of the past flybys of the Galileo, NEAR, and Rosetta spacecraft. An empirical formula proposed to explain the anomalies of such probes predicts an effect as large as about 6 mm s$^{-1}$  for the Juno's range-rate.

We looked at some dynamical effects which, in principle, may be considered as viable candidates   by numerically calculating their effects on the range-rate of Juno at its flyby of Earth. The Earth's quadrupole mass moment $J_2$, usually modeled in the data reduction softwares,  nominally shifts the Juno's range-rate by a few m s$^{-1}$ at the Newtonian level; by conservatively assuming an uncertainty in it of the order of $\approx 10^{-10}$ from the latest global gravity field models, the resulting residual signal reduces down to about 1 $\mu$m s$^{-1}$. The general relativistic Schwarzschild-type component of the Earth's gravitational field, which is modeled in the data analyses, causes a nominal range-rate shift of the order of 25  $\mu$m s$^{-1}$. The impact of the unmodeled general relativistic Lense-Thirring and $J_2 c^{-2}$ effects on the Juno's range-rate is at the $0.05-0.01$ $\mu$m s$^{-1}$ level. A putative Rindler-type radial uniform acceleration of the same magnitude as in the Pioneer anomaly would perturb the Juno's range-rate by 1  $\mu$m s$^{-1}$.

If a $\approx\ {\rm mm\ s^{-1}}$ flyby anomaly will finally result also for Juno, it will not be caused by any of the dynamical effects considered in this work. In particular, the unmodeled relativistic signatures will be too small to be detectable, regardless of any further consideration on the flyby anomaly as a sign of new physics.

\bibliography{flybybib}{}

\end{document}